**On estimating the size of overcoverage with the latent class model. A critique of the paper "Population Size Estimation Using Multiple Incomplete Lists with Overcoverage"** by di Cecco, di Zio, Filipponi and Rocchetti (2018, *Journal of Official Statistics* **34** 557–572)

Peter G.M. van der Heijden[1,2] and Paul A. Smith[1]

[1] S3RI and Dept of Social Statistics & Demography, University of Southampton, UK
[2] Utrecht University, The Netherlands

*Summary*

We read with interest the article by di Cecco et al. (2018), but have reservations about the usefulness of the latent class model specifically for estimating overcoverage. In particular, we question the interpretation of the parameters of the fitted latent class model.

*Introduction*

Di Cecco *et al.* (2018) propose to use the latent class model for estimating the size of the population. Using the latent class model, they have three aims: to account for overcoverage, dependence of captures in different lists and presence of incomplete lists. We focus here on their approach to take overcoverage into account, where they follow Biemer (2011, section 6.3). In the following, we expect that a reader has knowledge of their paper.

Our critique of their approach can be most easily understood from their simulation study (their Section 3). In their simulation study there are four registers, A, B, C and D, with indices a, b, c and d respectively. Indices a, b, c and d can take values 0 (absent) and 1 (present). They assume the existence of a dichotomous latent variable X with index x (x = 0, 1). The latent class model assumes that, given the latent variable X, the observed registers are statistically independent. In common loglinear model notation, the loglinear model in the unobserved table of five variables A, B, C, D and X is [AX][BX][CX][DX]. Notice that in the unobserved table for example A and B are independent conditional on X, but when we add up over X, A and B will generally be statistically dependent. Thus the latent variable "explains", one could say, the dependence between the registers.

We have one criticism and a remark. The first criticism is that we question the possibility Di Cecco et al. see to coin one latent class a latent class for overcoverage. The remark concerns the difficulty of proposing a correct latent class model.

*Criticism: in practice there will be no pure latent class with overcoverage*

In scenario 1 of their simulation study they use the following parameters of the latent class model (also with an interaction between registers C and D, but for the first point we want to make this is irrelevant): the total N = 1,000,000, the first latent class size is 0.4 and the second latent class size is 0.6. In other words, 40 % of the individuals fall in the first latent class and the 60% in the second latent class. The conditional parameters in the latent class model are

|               | Register A | Register B | Register C | Register D |
|---------------|------------|------------|------------|------------|
| Latent class 0 | 0.25       | 0.20       | 0.21       | 0.29       |
| Latent class 1 | 0.70       | 0.82       | 0.86       | 0.83       |

The latent class sizes are to be interpreted as follows: given that an individual is in Latent class 0, his/her probability to fall in register A is 0.25, in B 0.20, in C 0.21 and in D 0.29. In Latent class 1 these probabilities are much higher, 0.70, 0.82, 0.86 and 0.83.

*The standard interpretation* of the latent class model for population size estimation assumes that *there is no overcoverage*. The latent class model is then used in this context to allow for heterogeneity of inclusion probabilities. I.e. there are two groups of individuals, and one group of individuals (latent class 0) has low inclusion probabilities, namely 0.25, 0.20, 0.21 and 0.29; the other group (latent class 1) has high probabilities, namely 0.70, 0.82, 0.86 and 0.83. For latent class 0 there is a part of the population that is missed, and by estimating it using the conditional probabilities we also have an estimate for the population size $N_0$ for latent class 0. Similarly, for latent class 1 there is also a part of the population that is missed, and by estimating it using the conditional probabilities we also have an estimate for the population size $N_1$ for latent class 1. *The total population size N is then found by adding the estimates for $N_0$ and $N_1$.* In the context of no overcoverage but only undercoverage we have applied these models ourselves (Stanghellini and van der Heijden, 2004,, as was noticed by Di Cecco et al., 2018) and find the latent class model useful in the context of population size estimation.

*The Di Cecco et al. interpretation* of the latent class model assumes that *there is overcoverage.* They use the same parameter values but present them differently.

|               | Register A | Register B | Register C | Register D |
|---------------|------------|------------|------------|------------|
| Latent class 0 | 0.25       | 0.20       | 0.21       | 0.29       |
| Latent class 1 | 0.30       | 0.18       | 0.14       | 0.17       |

The parameters in the first line are *overcoverage* probabilities, i.e. in latent class 0, the probabilities of overcoverage are 0.25, 0.20, 0.21 and 0.29. In the second line of the second table they provide the complement of the probabilities in the second line of the first table and coin them undercoverage probabilities: given that one is in latent class 1, the inclusion probability is 0.70 so the undercoverage probability (the probability to be missed) is 0.30. The conclusion they draw from this interpretation is that *the population size N is only $N_1$, and $N_0$ should be ignored as it is overcoverage*.

Di Cecco et al. show in simulations that, if the scenario is true, the latent class model works well. What we question, however, is the likelihood that the scenario can be true in practice.

Our point is this. In practical situations there will be linked data of, for example, four registers. Usually it will be clear from the start whether one or more of the sources has overcoverage. If there is no overcoverage, however, then the Di Cecco et al. interpretation is wrong, and, where Di Cecco et al. would ignore $N_0$ in the estimation of the total population size N, they would do this incorrectly.

Assume that there is overcoverage. Then, however, it is likely that there will also be heterogeneity of inclusion probabilities. Thus, by fitting the 2-class latent class model the interpretation of latent class 0 is difficult as it will reflect both overcoverage *and* heterogeneity of inclusion probabilities. The problem then is that it is not clear what to do with the estimation of the total population size: part of $N_0$ should be ignored as it is overcoverage and as such *not* part of the population that we want to estimate the size of, but the part of $N_0$ resulting from heterogeneity should be included as it *is* part of the population that we want to estimate the size of. We think that in practical situations there will always be one or more subgroups of individuals for which the inclusion probabilities will be smaller than the inclusion probabilities of the main group – think of hard-to-reach groups, for example. Thus the model that they start off with in the simulations will never hold.

*Remark*

In a loglinear modelling approach where no use is made of a latent variable X, the loglinear model is likely to have interaction parameters that arise because of heterogeneity of inclusion probabilities. I.e., assume the simple two register situation where the model is [AX][BX], where X is a variable influencing the inclusion probabilities in A and B. Then, if the three way array is marginalized over X, interaction between A and B arises. In the latent class model this variable X is unobserved, and in the standard approach the latent variable X is used to eliminate direct dependencies between observed lists, so that the observed variables are independent conditional on the latent variable.

In the Di Cecco et al. interpretation, their initial model is a model of independence of the registers given the latent classes. To make the model more realistic, they allow for local dependencies, for example, in the first scenario of the simulation study in their paper, that also allows for direct interaction between registers C and D. Thus the model fitted is [AX][BX][CX][DX][CD]. They correctly notice that this assumes that the direct dependence between C and D is the same in latent class 0, the overcoverage class, and in latent class 1, the undercoverage class.

In the Di Cecco et al. interpretation the two latent classes actually refer to different populations: an overcoverage population of individuals that are not part of the population one is interested in, and an undercoverage population of individuals that is actually the population one is interested in. But actually this undercoverage population is a population from which the overcoverage is eliminated. But if we consider the undercoverage population separately, this is population of observed variables, that are likely to have direct interactions between them as these direct interactions arise from heterogeneous inclusion probabilities.

The problem we see is that there is no reason why the direct interactions needed to find a fitting model in the undercoverage class are identical to the direct interactions needed to find a fitting model in the overcoverage class. So, if we consider the direct interaction between C and D discussed above, it is not realistic to assume that the model [AX][BX][CX][DX][CD] would fit. Of course, this can be remedied by including a three-factor interaction between C, D and X so that the model becomes [AX][BX][CDX].

But then, what happens to the identifiability of the model if more interactions are to be included? This may easily become problematic. For four registers the number of degrees of freedom in the standard latent class model is (#independent cells - #independent parameters) = (15 – 1) – (1 + 4*2) = 5. For the model [AX][BX][CDX] it is 3, and by including additional interactions the number of degrees of freedom quickly becomes negative. This implies that, due to identifiability problems, it may be difficult to fit realistic models that fit the data.

*Conclusion*

Summarizing, our main critique is that, in the context of population size estimation using registers having overcoverage, it is very unlikely that there will not be heterogeneity of inclusion probabilities as well. This makes the interpretation of the parameters of the latent class model problematic, and ignoring $N_0$ for the estimation of the total population size problematic. Our conclusion is that we have strong hesitations about the usefulness of the latent class model in the context of overcoverage.